\documentstyle[12pt,prl,aps]{revtex}
\newcommand{\be}{\begin{equation}}
\newcommand{\ee}{\end{equation}}
\newcommand{\bay}{\begin{eqnarray}}
\newcommand{\eay}{\end{eqnarray}}
\newcommand{\conj}[1]{\overline{#1}}

\begin{document}
\input{psfig}
\begin{titlepage}
\begin{flushright}
IFUP-TH 1/96\\
January 1996\\
quant-ph/9601024
\end{flushright}

\vspace{5mm}

\begin{center}

{\Large\bf The interaction time of a packet with a

\vspace{3mm}

potential barrier}

\vspace{12mm}
{\large Andrea Begliuomini$^{a}$ and Luciano Bracci$^{a,b}$\\}

\vspace{3mm}

$(a)$ Dipartimento di Fisica dell'Universit\`{a},\\
piazza Torricelli 2, I-56126 Pisa, Italy.\\
$(b)$ Istituto Nazionale di Fisica Nucleare, sezione di Pisa, Italy.\\
\end{center}

\vspace{2truecm}

\begin{quote}
\centerline{\bf Abstract}
We study the evolution of a wave packet impinging onto a one dimensional
potential barrier.
The transmission and reflection times discussed in the literature for
stationary states do not correspond to the times required for the emergence
of a transmitted or a reflected packet.
We propose new definitions for the interaction (dwell) time and the
transmission and reflection times which are suitable for packets and fit
better the actual time evolution of the packet.

\end{quote}

\vspace{0.5truecm}
PACS numbers: 03.65.Bz,73.40.Gk

\end{titlepage}
\clearpage

\section{Introduction}
The tunneling of a particle beyond a potential barrier is one of the
simplest effects predicted by Quantum Mechanics where the conflict
between classical and quantum pictures is most striking. Although it has
been studied since the early days of Quantum Mechanics \cite{maccoll},
the debate is still open as to what should be meant for tunneling time
(see refs. \cite{hast89}-\cite{lama94} for extensive reviews). 

It could be objected that in a proper quantum formulation of the
problem there is no room for such a concept, all that we can ask being the
probability of detecting the particle beyond the barrier or of having it
reflected by the potential. We feel, however, that this is an extreme view.
It is legitimate, for example, to think of an ensemble of systems prepared
in a given initial state, for which the times are measured when a detector
located in front or beyond the barrier reveals the arrival of the particle
(in order not to disturb the state, we think of separate experiments for
measuring the arrival and the exit times). Knowledge of the initial state
(i.e. the wave function at time $t=0$) should enable us to determine which is
the difference between the average exit time and the average arrival time.
This difference could be regarded as the average tunneling time. While we
do not advocate the above definition as the definition of the tunneling
time, we insist that the wave function contains implicitly the information
as to the time that in the average the particle spends in the potential
barrier. The problem with the determination of this average time is that,
the process being intrinsically non-classical, it is not possible to look for
a quantum counterpart of a classical observable, whose average value on
the state of the particle should be interpreted as the tunneling time. On
the other hand, a clearer view about the tunneling time is  urged also by
current experiments on semi-conductor devices \cite{gueret} and evanescent
waves \cite{evane}, where observation of superluminal velocity has sometimes
been claimed. 

Most of the approaches to the problem of tunneling time deal with
stationary states. The particle is in an eigenstate of the Hamiltonian, and
the tunneling time $\tau_{\rm T}$, as well as the reflection time
$\tau_{\rm R}$ and
the dwell time $\tau_{\rm D}$, are functions of its energy $E$, or momentum
$k=\sqrt{2mE/\hbar}$.
Also those who envisage wave packets (see \cite{olre92}, \cite{hart}, \cite{stev})
use packets which are so narrow in energy as to allow the monochromatic
approximation to hold.
We have preferred to investigate the time evolution of a wave packet with a
spread in energy which forbids this approximation.
The packet impinges onto the barrier and is partially reflected and partially
transmitted.
It is constructed as a Gaussian superposition of eigenstates centered around a
value $E$ lower than the height $V_0$ of the barrier.
We observe the evolution of the packet at times $\tau_{\rm T}(E)$, $\tau_{\rm R} (E)$ and
$\tau_{\rm D}(E)$ after it has reached the barrier.
Although the time $t_{\rm in}$ when the packet "reaches the barrier" is not
sharply defined, due to the interference of the higher energy components,
which reach the barrier earlier and are partially reflected, with the lower
energy incoming components, we find that the above mentioned times
$\tau_{\rm T}(E)$, $\tau_{\rm R}(E)$ and $\tau_{\rm D}(E)$ definitely do not correspond to the times required for the packet to emerge from the barrier.
The uncertainty in $t_{\rm in}$ is not such as to alter this conclusion.
It can be argued that the transmitted packet is centered around an energy
$E_{\rm T}$ higher than $E$, whereas the reflected packet is centered around
an energy $E_{\rm R}$ lower than $E$, but also the times
$\tau_{\rm T}(E_{\rm T})$
or $\tau_{\rm R}(E_{\rm R})$ are by no means representative of the times required to see the reflected packet or the transmitted packet.
The conclusion seems to be that the tunneling times defined for stationary
states with energy $E$ are not meaningful for the evolution of a wave packet
having $E$ as average energy.

This has prompted us to look for another determination of the times
$\tau_{\rm T}$, $\tau_{\rm D}$ and $\tau_{\rm R}$  which take into account
the actual behaviour of the wave packet.
As for $\tau_{\rm D}$, we define it as the time integral of the probability
$P_2(t)$ of finding the particle within the barrier region,
\be
\tau_{\rm D}  =  \int_{t_2}^{+\infty} P_2(t) dt
\label{tau_D}
\ee
this definition being unsensitive to the time $t_2$ when the particle begins
to interact with the barrier, provided it is earlier than the time when the
packet impinges onto the barrier. Thus, $\tau_{\rm D}$ can be regarded as the total
interaction time. As for $\tau_{\rm T}$, we define it as a weighted sum of every time
interval $\Delta t$, the weight being the fraction of the transmitted packet
which at time $t$ has not yet been transmitted.
The definition for $\tau_{\rm R} $  is similar. 
The trouble with these definitions is that there is still a problem with the
time when the integral over $t$ begins.
In principle, the packet is interacting with the barrier since time $t=0$,
but  the contribution of the earlier times, when the particle has not yet
arrived onto the barrier, to $\tau_{\rm T}$ should be negligible.  

While there is no objective way of determining the time when the
interaction between the particle and the potential starts, we observe that
a shifting of $t_2$ in eq. (\ref{tau_D}) affects $\tau_{\rm D}$, whose value decreases
with increasing $t_2$.
On the other hand any determination of $\tau_{\rm D}$ entails a given accuracy
$\epsilon$.
For the lower integration limit in the definitions of $\tau_{\rm T}$ and $\tau_{\rm R} $
we choose the value $t_{\rm L}$ such that $\int_{t_2}^{t_{\rm L}} P_2(t) dt$
equals $\epsilon$.
The meaning of this choice is that we have a finite time resolution $\epsilon$,
and we neglect those time intervals which contribute to the interaction time
less than $\epsilon$. We note however that the time dependence of the
probability $P_2(t)$ and of the analogous probabilities of finding the
particle beyond the barrier ($P_3(t)$) and in front of it ($P_1(t)$) when the
packet impinges onto the barrier is sufficiently steep so that the choice of
$t_{\rm L}$ does not really affect $\tau_{\rm T}$ and $\tau_{\rm R}$ in a substantial way.
The values we find (with $\epsilon \simeq 0.01$) for $\tau_{\rm T}$,
$\tau_{\rm R}$ and $\tau_{\rm D}$ are definitely different from the values of
$\tau_{\rm T}(E)$, $\tau_{\rm R}(E)$ and $\tau_{\rm D}(E)$ and are in agreement
with the actual time evolution of the packet. 

We have also examined the behaviour of the probabilities $P_i(t)$ for large
$t$. These probabilities have an exponential tail $e^{-t/\tau}$, with the same
time constant $\tau$ (depletion time) for each probability. The value of
$\tau$ is determined by the position of the poles of the transmission
coefficient $D(k)$ in the complex $k$ plane, and is independent of the
details of the wave packet.

In conclusion, we find that the tunneling times proposed in the
literature for stationary states are not meaningful for the actual time
evolution of a wave packet. The time lapse from the first contact of the
packet with the barrier to its emergence beyond the barrier, although not
so sharply definable, is definitely different from the tunneling times
proposed for stationary states. 

In section 2 we define the problem and build the packet whose time
evolution is discussed in section 3. In section 4 we present the definitions
of $\tau_{\rm D}$, $\tau_{\rm T}$ and $\tau_{\rm R} $ and in section 5 we discuss the depletion
time. Section 6 is devoted to the conclusions. 

\section{The wave packet}

We consider a one dimensional problem, the Hamiltonian being

\be
H=\frac{p^2}{2m}+V(x)
\label{hamiltonian}
\ee
\be  
V(x)=
\left \{ \begin{array}{ll}
\displaystyle{V_0 \qquad} & ~(|x|<d) \\
\displaystyle{0 \qquad} & ~(|x|>d)
\end{array} \right.
\label{potential}
\ee
We use units such that $\hbar=1$. The eigenfunctions $\psi_k$ of the
Hamiltonian (\ref{hamiltonian}) are well known (see Appendix).
We consider the evolution of a wave packet $\psi$ built as a Gaussian
superposition of the functions $\psi_k$, impinging onto the
barrier:
\be
\psi(x,t) = \int a(k) \, \psi_k(x) \, e^{-ik^2t/2m} \, dk 
\label{psi}
\ee
where the coefficients $a(k)$ are
\be
a(k)=\left(\frac{2 \delta^2}{4 \pi^3}\right)^{1/4} e^{-(k-k_{\rm av})^2}
e^{-ikx_0}
\label{ak}
\ee
$k_{\rm av}$ is the average momentum, and $x_0$ is the coordinate of the peak
of the packet at $t=0$.
We choose the parameters involved in the problem as follows:
\be
m=1 \qquad k_{\rm av}=9.9 \qquad \delta=\sqrt{2} \qquad x_0=-15 \qquad
d=2 \qquad V_0 =50\,\,(k_0=10)
\label{parameters}
\ee
The packet (\ref{psi}) is Gaussian also in $x$ (fig. 1).
The reason why the peak of the packet is located so far from the left edge
of the barrier at time $t=0$ is to have an identification of $t_{\rm in}$
as sharp as possible.
With a packet starting nearer to the barrier, the Gaussian form
of the packet would be immediately lost, due to the interference of the
incoming and reflected components of the packet.

\section{Evolution of the packet}

We have studied the evolution of the packet in order to verify to what
extent the definitions of the tunneling time proposed for stationary
problems are meaningful for a wave packet. More precisely, we have tested
whether the phase times $\tau^{\rm ph}$ \cite{hart} or the times proposed
by Buttiker $\tau^{\rm B}$ \cite{bu83}
(which turn out to be deeply connected with the times proposed within
other approaches, see for \cite{hast89}-\cite{lama94} a review) do represent
the lapse of time which the packet spends in the barrier.
To this purpose, it is necessary to mark the time $t_{\rm in}$ when the packet
begins to interact with the barrier, and the time $t_{\rm fin}$ when a
transmitted (reflected) packet appears.
The comparison of $t_{\rm fin} - t_{\rm in}$ with $\tau^{\rm ph}$ and
$\tau^{\rm B}$ calculated for significant values of the energy will show that
$\tau^{\rm ph}$ and $\tau^{\rm B}$ are not significant for a wave packet.

We observe the shape of the packet as it moves towards the potential
barrier. As long as it is far enough, its shape is quite similar to the initial
shape; when it approaches the left edge of the barrier it begins to become
blurred (fig. 2) due to the interference between the incoming and the
reflected components. There is a time interval in which the interference
phenomenon is dominant, but a reflected packet is still absent. The peak of
a reflected packet appears a time $\Delta t$ after the blurring of the incoming
packet. A bit later, we see the emergence of a transmitted packet beyond
the barrier. 

For the time $t_{\rm in}$ when the packet begins to interact with the barrier
we assume the time when the blurred shape can be macroscopically observed.
As for the time $t_{\rm fin}$, we have two possibilities: the time when the
reflected packet appears ($t_{\rm fin,R}$) or the time when the peak of the
transmitted packet appears ($t_{\rm fin,T}$).
It is clear from the above that neither $t_{\rm in}$ nor $t_{\rm fin,R}$
($t_{\rm fin,T}$) are sharply defined. In the units we have chosen,
each of them can be determined only within an error $\Delta t \simeq 0.1$. 

By inspection of the graphs representing the reflected and the
transmitted packet respectively (see figs. 2a and fig2b),
we get
\be
t_{\rm in} \simeq 0.9 \qquad t_{\rm fin,R} \simeq 1.9 \qquad
t_{\rm fin, T}=2.7
\label{inspection}
\ee
For the  times $\tau_{\rm R}$  and $\tau_{\rm T}$ it follows
\be
\tau_{\rm R}  \simeq 1, \, \tau_{\rm T} \simeq 1.8
\label{rough}
\ee
The error on the above values can be assessed to be of order 0.2.

We compare the above times with the times $\tau^{\rm ph}$ and $\tau^{\rm B}$.
These are functions of the momentum $k$, so we must decide which momentum to
consider. We consider the average momentum $k_{\rm av}$ and the average momenta
$k_{\rm R} $ and $k_{\rm T}$ of the reflected and transmitted components respectively. These
latter are calculated to be
\be
k_{\rm R} = 9.696, \, k_{\rm T} = 10.327
\label{kparticular}
\ee
Incidentally, this shows, as previously noted \cite{lama94}, that a potential
barrier acts as an accelerator: the transmitted wave packet has an average
momentum larger than the incoming one. The opposite holds for the
reflected packet.

In table 1 we present the dwell times $t^{\rm ph,D}$ and
$t^{\rm B,D}$ calculated for $k=k_{\rm av}$, $k_{\rm R} $ and $k_{\rm T}$,
together with the reflection time $t^{\rm B,R}$ calculated for $k_{\rm av}$ 
and $k_{\rm R}$ and the transmission time $t^{\rm B,T}$ calculated for $k_{\rm av}$ and $k_{\rm T}$. We see
that the dwell times $t^{\rm ph,D}$ and
$t^{\rm B,D}$ are definitely shorter than the
tunneling time $\tau_{\rm T}$ reported in (\ref{rough}).
The same holds for the reflection times
$t^{\rm B,R}$ calculated for $k_{\rm av}$ and $k_{\rm R}$.
As for the transmission time $t^{\rm B,T}$, we see
that the value corresponding to $k_{\rm av}$ is longer, whereas the value
corresponding to $k_{\rm T}$ is shorter. We conclude that the tunneling times
found for stationary problems are not useful to describe the evolution of a
packet.

\section{Definition of $\tau_{\rm D}$, $\tau_{\rm T}$ and $\tau_{\rm R}$}

So far definitions of the time that the particle interacts with the
potential (the so called dwell time $\tau_{\rm D}$) and of the transmission
($\tau_{\rm T}$) and reflection ($\tau_{\rm R}$) times have been given mainly
for stationary problems.
Even the authors who have dealt with wave packets considered packets which
were so narrow in energy that the relevant times could be considered to be
functions $\tau(k)$ of the momentum, as in the stationary case. In this section
we propose definitions of $\tau_{\rm D}$, $\tau_{\rm T}$ and $\tau_{\rm R}$
which are suitable for a wave packet.

We first consider the probability $P_2(t)$ that at time $t$ the particle is
within the barrier region,
\be
P_2(t)=\int_{-d}^{d} dx\,\left|\psi(x,t)\right|^2
\label{p2}
\ee
and the analogous probabilities that at time $t$ the particle is in front of
the barrier ($P_1(t)$) or beyond the barrier ($P_3(t)$).
Obviously, we have $P_1(t) + P_2(t) + P_3(t) = 1$ 

The values of $P_1(t)$, $P_2(t)$ and $P_3(t)$ are reported in fig. 4.
$P_1$ and $P_3$ tend to asymptotic values which we call respectively R and T.
They represent the probabilities that the particle is reflected or transmitted
respectively.
Obviously, R + T = 1.

Inspection of $P_2(t)$ shows that for $t \leq 0.75$ the packet does not
interact with the barrier, and the interaction reaches its maximum at
$t \simeq 1.5$.
After this time the probability of finding the particle in the barrier region
decreases with a tail which has an exponential shape. The interaction time
should be the total time that the particle spends in the potential region.
With this view, we propose the following (already deined as eq. (\ref{tau_D}))
as a definition of $\tau_{\rm D}$
(see also ref. \cite{muga}, \cite{leae93} where a similar definition is
put forth):
$$\tau_{\rm D}  =  \int_{t_2}^{+\infty} P_2(t) dt$$

The meaning of eq. (\ref{tau_D}) is clear: every time interval $dt$ is weighted
with the probability $P_2(t)$ of finding the particle within the potential
barrier. The dwell time is to be interpreted as the time the particle
interacts with the potential regardless its fate.
The definition is independent of the choice of time $t_2$, provided it is
chosen earlier than the time the packet is significantly present in the
barrier region. We can safely take $t_2=0$.

We define the transmission time $\tau_{\rm T}$ as the average time that it
takes for the transmitted particles to emerge beyond the potential barrier.
This time is given by the integral
\be
I = \int_{t_3}^{\infty} \left(1 - \frac{P_3(t)}{T}\right) dt
\label{integral}
\ee
In the above definition $1 - P_3(t)/T$ is the fraction of the transmitted
packet which has not yet been transmitted at time $t$. We interpret this
fact viewing this fraction as "being transmitted" at time $t$, so that any
time interval $dt$ contributes to the transmission time with a weight
$1 -  P_3(t)/T$.

The trouble with this definition is the lower integration limit $t_3$, in
that eq. (\ref{integral}) gives weight 1 also to the time intervals when the
packet has not yet arrived in the potential region.
But these time intervals are not to
be considered as contributing to the transmission time: what eq. (\ref{integral})
actually gives is the transmission time starting from time $t_3$. 

The problem cannot be circumvented by any choice of time $t_3$. In
principle, $t_3$ should be the time when the particle begins to interact with
the potential, but this time cannot be determined in any objective way.
However, we can determine that time $t_\epsilon$ such that the packet has
spent in the potential region an amount $\epsilon$ of time starting from the time $t=0$. This
is the time $t_\epsilon$ such that the integral of $P_2$ from $t=0$ to
$t=t_\epsilon$ equals $\epsilon$. Now,
the transmission time $\tau_{\rm T}(\epsilon)$ reckoned from time $t_\epsilon$
(such that the time spent in the barrier by the packet is equal to $\epsilon$)
is defined unambiguously.
On the other hand, the calculation of the dwell time $\tau_{\rm D}$
(as well as any possible time measurement about the particle) is affected by
an error. If we choose $\epsilon$ to be the same as this error, the uncertainty
on $\tau_{\rm T}$ due to the choice of the lower integration limit can be
considered of the same order as the time resolution we are able to attain.
In conclusion, we put
\be
\tau_{\rm T}(\epsilon) = \int^{\infty} \Theta\left(\int^t P_2(x)dx - \epsilon
\right)\left[1-\frac{P_3(t)}{T}\right] \, dt
\label {tau_T_epsilon}
\ee
where $\Theta(x)$ is the Heavyside step function.
The lower integration limit in eq. (\ref{tau_T_epsilon}) can be taken the same
as in eq. (\ref{integral}).

Along these lines we define also the reflection time $\tau_{\rm R}(\epsilon)$:
\be
\tau_{\rm R}(\epsilon) = \int^{+infty} \Theta\left(\int^t P_2(x)dx - \epsilon
\right)\left[1-\frac{P_1(t)}{R}\right] \, dt
\label {tau_R_epsilon}
\ee

The interpretation of eq. (\ref{tau_R_epsilon}) is straightforward.
$1-P_1(t)/R$ is the fraction of the reflected packet that at time t has not
yet been reflected. This fraction is taken as the weight for any time interval
$dt$. We note however that for times near the beginning of the interaction of
the particle with the potential this weight can be negative. But the decrease
of $P_1(t)$ is very sharp (see fig. 4) and the contribution to the
integral of the region where the weight is negative is small indeed.

In the case we have investigated we have found
\be
\tau_{\rm D} = 0.93
\label{tau_D_num}
\ee
In order to find $\tau_{\rm T}$ and $\tau_{\rm R}$  we have fixed
$\epsilon=10^{-2} \tau_{\rm D} \simeq 0.01$. This yields
\be
\tau_{\rm T} = 3.39 \qquad \tau_{\rm R}  = 0.55
\label{times}
\ee
Note that we have $T \simeq 0.14$, $R \simeq 0.86$.
With this values and the results reported in eqs. 14 and 15 the conditional
probabilty relation (\cite{hast89}, \cite{leae93}
\be
\tau_{\rm D} = T\tau_{\rm T} + R\tau_{\rm R}
\label{condition}
\ee
is fairly satisfied. Condition (\ref{condition}) would be identically
satisfied if the lower limits where the integrands in eqs. (\ref{tau_D}),
(\ref{tau_T_epsilon}) and (\ref{tau_R_epsilon}) start to differ
from zero were the same. The fact that eq. (\ref{condition}) holds true with a
fair accuracy, to within that value .01 which can be assessed as the accuracy
of all our calculations, can be regarded as a support to the correctness of
the definitions of $\tau_{\rm D}$, $\tau_{\rm T}$ and $\tau_{\rm R}$.

By inspecting table 1 we see that $\tau_{\rm D}$ is definitely larger
than $\tau_{\rm ph,D}$ and $\tau_{\rm B,D}$ evaluated for $k=k_{\rm av}$.
The dwell time $\tau_{\rm D}$ looks a very reliable estimate
of the interaction time. The discrepancy with the value found for
stationary states confirms that the extrapolation to wave packet is
untenable. As for $\tau_{\rm T}$ and $\tau_{\rm R}$, a comparison with the
previously reported values of the reflection and transmission times derived by
inspection of the wave packet evolution shows that $\tau_{\rm T}$ and
$\tau_{\rm R}$  in eq. (\ref{times}) are respectively
longer and shorter. Thus, the discrepancy between the times in eq. (\ref{times})
and the times derived within the stationary approach, for example the Buttiker
times calculated for $k=k_{\rm av}$,  is even larger. 

\section{The depletion rate}

The tail of $P_2(t)$ can be described as an exponential curve with a time
constant $\tau_{\rm dep}$: 
\be
P_2(t) = A e^{-t/\tau_{\rm dep}} \qquad {\rm for \,\, large \,\,} t
\label{p2_as}
\ee
By considering the values of $P_2$ for $t>30$, we find that the exponential fit
is excellent, with 
\be
\tau_{\rm dep} = 16.192
\label{tau_dep}
\ee
and a correlation coefficient $R=-0.9999883$. The same time constant
$\tau_{\rm dep}$ rules the asymptotic behaviour of $P_3(t)$ and, due to
probability conservation, of $P_1$: 
$$\tau_{\rm dep}=16.192 \qquad R =  -0.9999435$$
The value of $\tau_{\rm dep}$ is connected with the behaviour of the complex
transmission coefficient $D(k)$ in the complex plane. By explicitly writing
$P_2(t)$ (see Appendix) we see that the only singularities are in the product
$u(k)u^*(p)$ in the denominator, with
\be
u(k) = (\kappa^2-k^2)\sinh(2\kappa d) - 2ik\kappa \cosh(2\kappa d)
\label{u}
\ee
It is easy to see that if $u(k)=0$, then $u^*(k^*)=0$. Hence, a zero for $u$ in
$k=x+iy$, together with the zero in $x-iy$ for $u^*$, will contribute to
$P_2(t)$ with a term $exp(2xyt/m)$. For large values of $t$, the term with $xy$
negative and minimum in absolute value will dominate.
The behaviour of $P_2$ will be as in eq. (\ref{p2_as}), with:
\be
\tau_{\rm dep}=m/2|xy|
\label{p2_ana}
\ee
The search for the pole with smallest $|xy|$ value is discussed in the
appendix. We find $x=10.03$, $y=-3.0565 \cdot 10^{-3}$, which yields
$\tau_{\rm dep}=16.3087$,
in fair agreement with the value reported in eq. (\ref{tau_dep}).

\section{Conclusions} 

We have studied the evolution of a wave packet which approaches a
potential barrier. The time requested for the appearance of a transmitted
($\tau_{\rm T}$) or a reflected ($\tau_{\rm R}$) packet, reckoned from the
moment the incoming packet begins to interact with the barrier, are definitely
different from the values of $\tau_{\rm T}$ and $\tau_{\rm R}$ reported in the
literature for monochromatic packets. We propose new definitions of
$\tau_{\rm D}$, $\tau_{\rm T}$ and $\tau_{\rm R}$ which are suitable for
a packet and give results that fit better the actual time evolution of the
packet.

The probability of finding the particle within the barrier has an
exponential tail whose time constant $\tau_{\rm dep}$ is determined by the
behaviour of the stationary solutions in the complex momentum plane.

\appendix
\section*{}

The stationary solutions of the Schroedinger equation with the
Hamiltonian (\ref{hamiltonian}) are:
$$\psi_k(x)=
\left \{ \begin{array}{ll}
\displaystyle{e^{ikx}+A(k)e^{-ikx}} \qquad {\rm if} \qquad x<-d \\
\displaystyle{B(k)e^{\kappa x}+C(k)e^{-\kappa x}} \qquad {\rm if} \qquad
|x|<d \\
\displaystyle{D(k)e^{ikx}} \qquad {\rm if} \qquad x>d
\end{array}
\right.$$
where 
$$A(k)=-\frac{(\kappa^2+k^2)\sinh(2\kappa d)}{u(k)}e^{-2ikd}$$
$$B(k)=-\frac{ik(\kappa+ik)}{u(k)}e^{-ikd}e^{-\kappa d}$$
$$C(k)=-\frac{ik(\kappa-ik)}{u(k)}e^{-ikd}e^{\kappa d}$$
$$D(k)=-\frac{2ihk}{u(k)}e^{-2ikd}$$
$$k=\sqrt{k_0^2-k^2}$$
$$u(k) = (\kappa^2-k^2)\sinh(2\kappa d) - 2ik\kappa \cosh(2\kappa d)$$
The probability $P_2(t)$ is given by:
\bay
&&P_2(t)=\sqrt{\frac{\delta^2}{2 \pi^3}}\int dk \int dp\,
e^{-(p-k_{\rm av})^2\delta^2} e^{-(k-k_{\rm av})^2\delta^2} e^{-i(k^2-p^2)t/2m}
\cdot \nonumber \\
&&\frac{kpe^{i(p-k)d}}{\left[\left(q^2-p^2\right)\sinh(2qd)-2ipq\cosh(2qd)
\right]^*\left[\left(\kappa^2-k^2\right)\sinh(2\kappa d)-2ik\kappa
\cosh(2\kappa d)\right]} \cdot \nonumber\\
&&\left[(q+ip)^*(\kappa+ik)\frac{\left(1-e^{-2(\kappa+\conj{q})d}\right)}
{\kappa+\conj{q}}+(q-ip)^*(\kappa+ik)\frac{\left(1-e^{-2(\kappa-\conj{q})d}
\right)}{\kappa-\conj{q}}+\right. \nonumber \\
&&\left.-(q+ip)^*(\kappa-ik)\frac{\left(1-e^{2(\kappa-\conj{q})d}\right)}
{\kappa-\conj{q}}-(q-ip)^*(\kappa-ik)\frac{\left(1-e^{2(\kappa+\conj{q})d}
\right)}{\kappa+\conj{q}}\right] \nonumber
\eay
where $q=\sqrt{k_0^2-p^2}$

The only singularities are in the factors $u(k) u^*(p)$ in the denominator.
In order to find the zeroes of $u(k)$ we note that if $u(k)=0$, then
\be
\left[\frac{\sin\left(2ik_0\sqrt{1-z^2}\right)}
{\sqrt{1-z^2}}\right]^2=4z^2
\label{ucondition} 
\ee
with $z=k/k_0$. For an opaque barrier (i.e. $2k_0d \gg 1$) the solution for $z$ has
to be near $z=1$. The values of the zeroes have been found by solving for $z$
the relation
$$0= u(z) \simeq u\left(z_0\right)+\left(z-z_0\right) u^{\prime}\left(z_0\right)$$ 
and iterating, choosing different values of $z_0$ near 1. We have found as
many different solutions as predicted by the Cernlib routine nzeros for a
neighbourhood of $z=1$. The value of $\tau_{\rm dep}$ is given by that
solution for which $xy$ is negative and minimum in absolute value.

%%%%%%%%%%%%%%%%%%%%%%%%%%%%%%%%%%%%%%%%%%%%%%%%%%%%%%%%%%%%%%%%%%%%%%%%%%%%%%%
\newpage
\centerline{\bf FIGURE CAPTIONS}
%%%%%%%%%%%%%%%%%%%%%%%%%%%%%%%%%%%%%%%%%%%%%%%%%%%%%%%%%%%%%%%%%%%%%%%%%%%%%%%
%Caption fig.1
{\noindent}{\bf Figure 1}\\
The Gaussian shape of the incoming packet for $t=0$.\\
\\
\\
%%%%%%%%%%%%%%%%%%%%%%%%%%%%%%%%%%%%%%%%%%%%%%%%%%%%%%%%%%%%%%%%%%%%%%%%%%%%%%%
%Caption fig. 2a and 2b
{\noindent}{\bf Figure 2a and 2b}\\
In (a) the wave packet at $t=0.9$ and in (b) a zoom of the blurred region.\\
\\
\\
%%%%%%%%%%%%%%%%%%%%%%%%%%%%%%%%%%%%%%%%%%%%%%%%%%%%%%%%%%%%%%%%%%%%%%%%%%%%%%%
%Caption fig.3a and 3b
{\noindent}{\bf Figure 3a and 3b}\\
In (a) the reflected packet that appears at $t=1.9$. In (b) the first
transmitted peak that appears at $t=2.7$.\\
\\
\\
%%%%%%%%%%%%%%%%%%%%%%%%%%%%%%%%%%%%%%%%%%%%%%%%%%%%%%%%%%%%%%%%%%%%%%%%%%%%%%%
%Caption fig.4
{\noindent}{\bf Figure 4}\\
Plots of $P_1(t)$, $P_2(t)$ and $P_3(t)$ as functions of $t$.\\
\\
\\
\\
\\
\\
\\
%%%%%%%%%%%%%%%%%%%%%%%%%%%%%%%%%%%%%%%%%%%%%%%%%%%%%%%%%%%%%%%%%%%%%%%%%%%%%%%
\centerline{\bf TABLE CAPTIONS}
{\bf Table 1}\\
The phase time and the B\"uttiker times evaluated in $k=\overline{k},
k_R, k_T$.
%%%%%%%%%%%%%%%%%%%%%%%%%%%%%%%%%%%%%%%%%%%%%%%%%%%%%%%%%%%%%%%%%%%%%%%%%%%%%%%
\newpage
\begin{center}
\begin{tabular}{|c|c|c|c|}
\hline
time & $k=k_{\rm av}$ & $k=k_R$ & $k=k_T$ \\
\hline
$\tau^{\rm ph,D}$ & 0.143 & 0.0843 & 1.011 \\
\hline
$\tau^{\rm B,D}$ & 0.140 & 0.079 & 1.008 \\
\hline
$\tau^{\rm B,T}$ & 2.357 & - & 1.248 \\
\hline
$\tau_{R}^{\rm Bu}$ & 0.140 & 0.079 & - \\
\hline
\end{tabular}
\end{center}
\centerline{\bf Table 1}
%%%%%%%%%%%%%%%%%%%%%%%%%%%%%%%%%%%%%%%%%%%%%%%%%%%%%%%%%%%%%%%%%%%%%%%%%%%%%%%
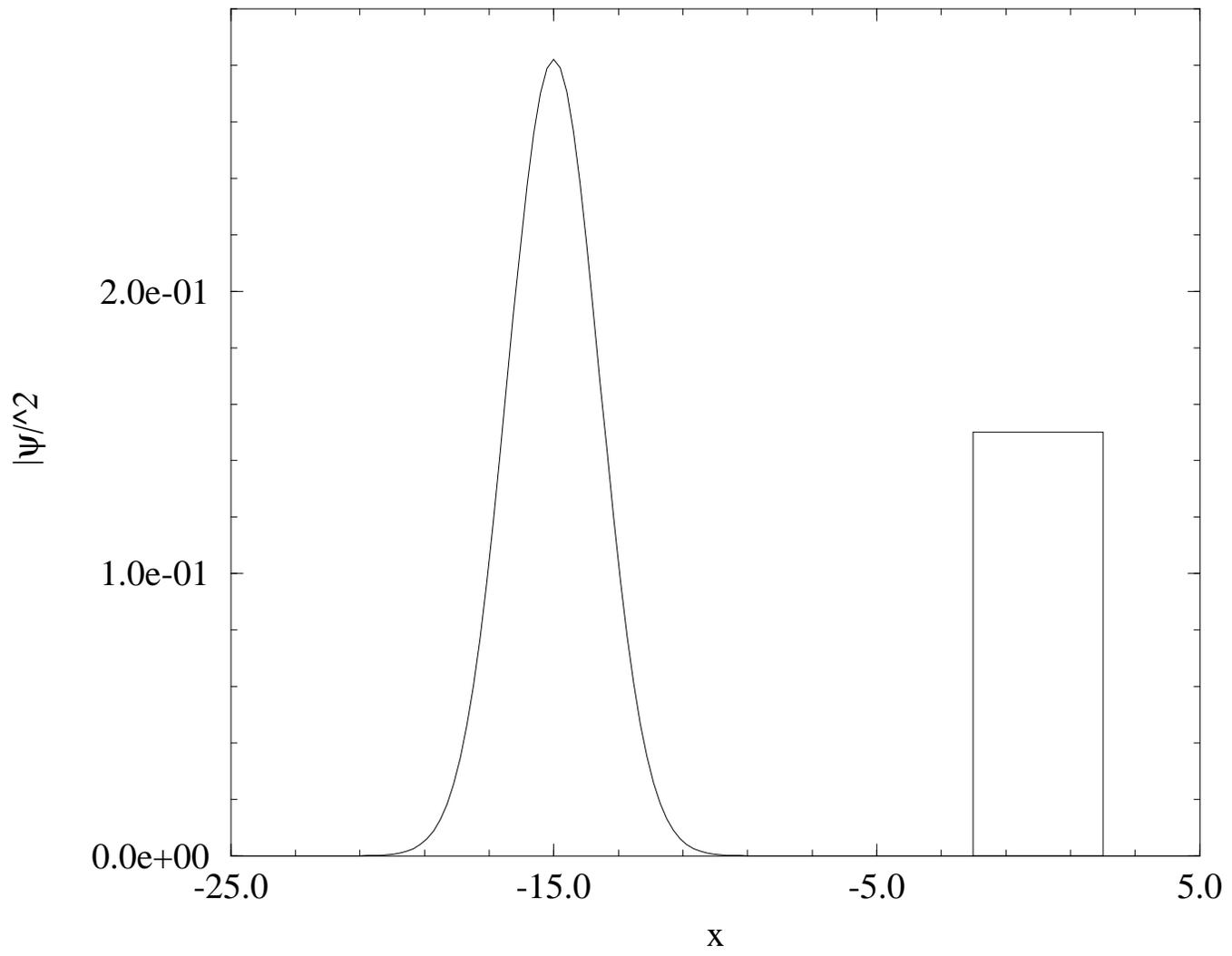
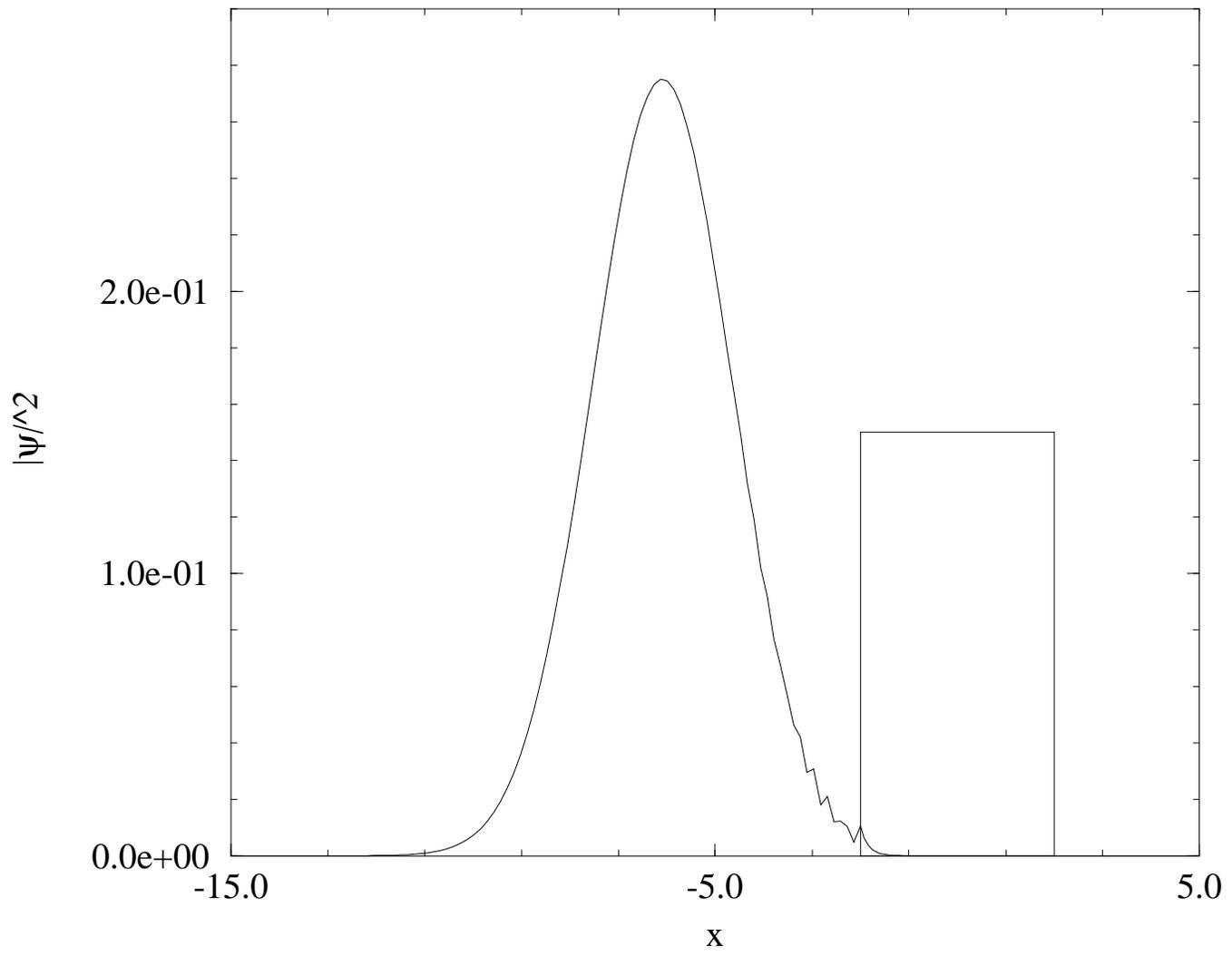
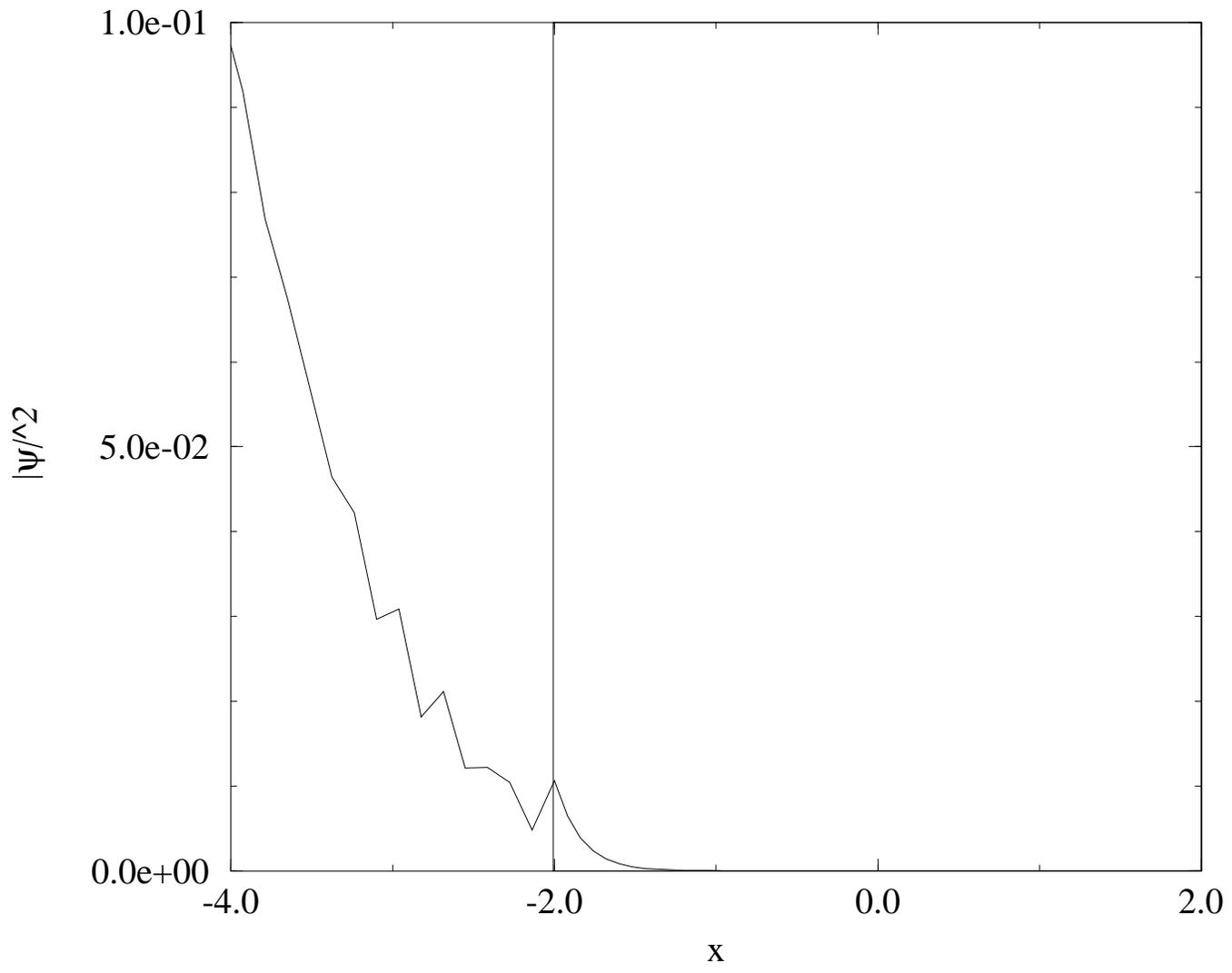
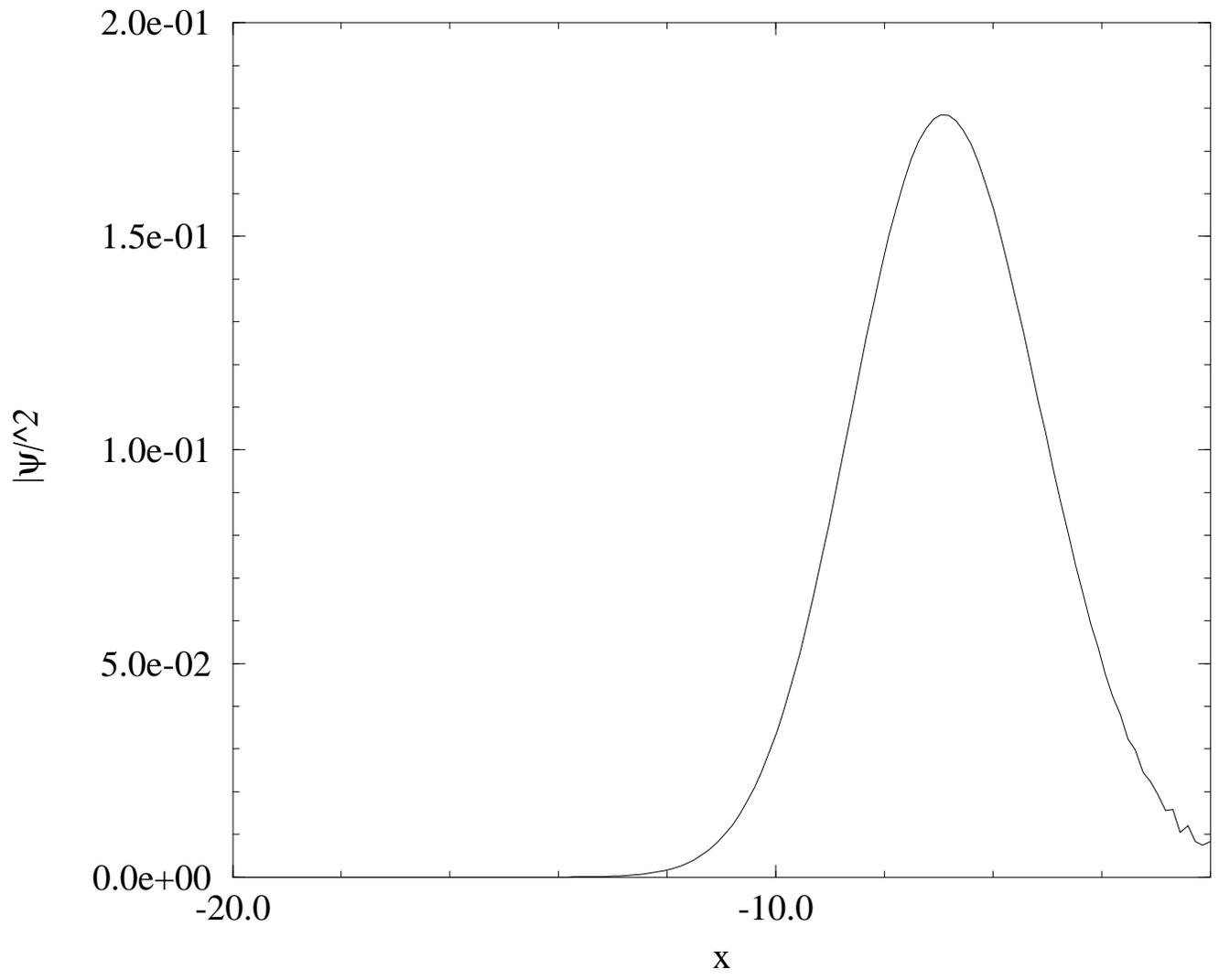
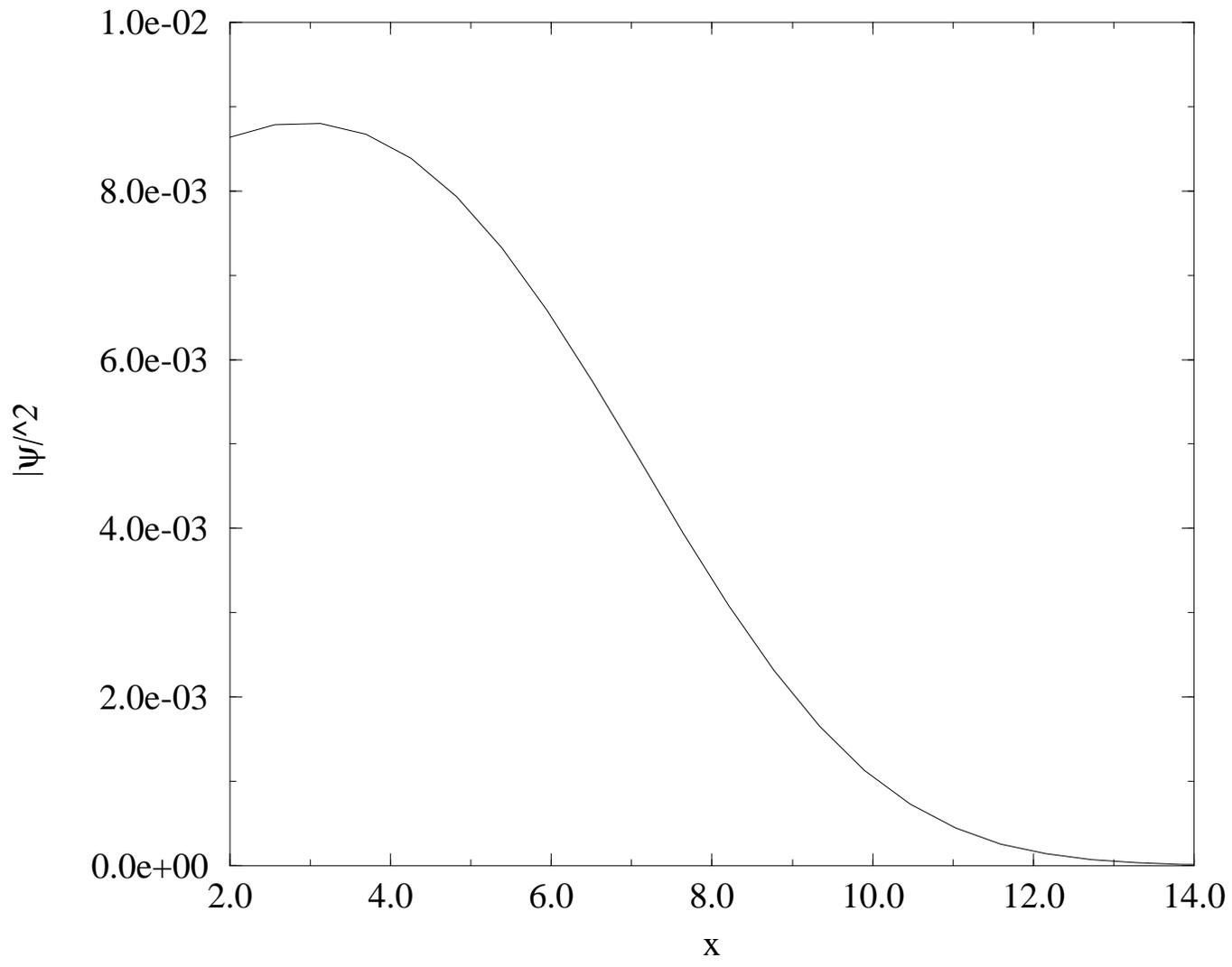
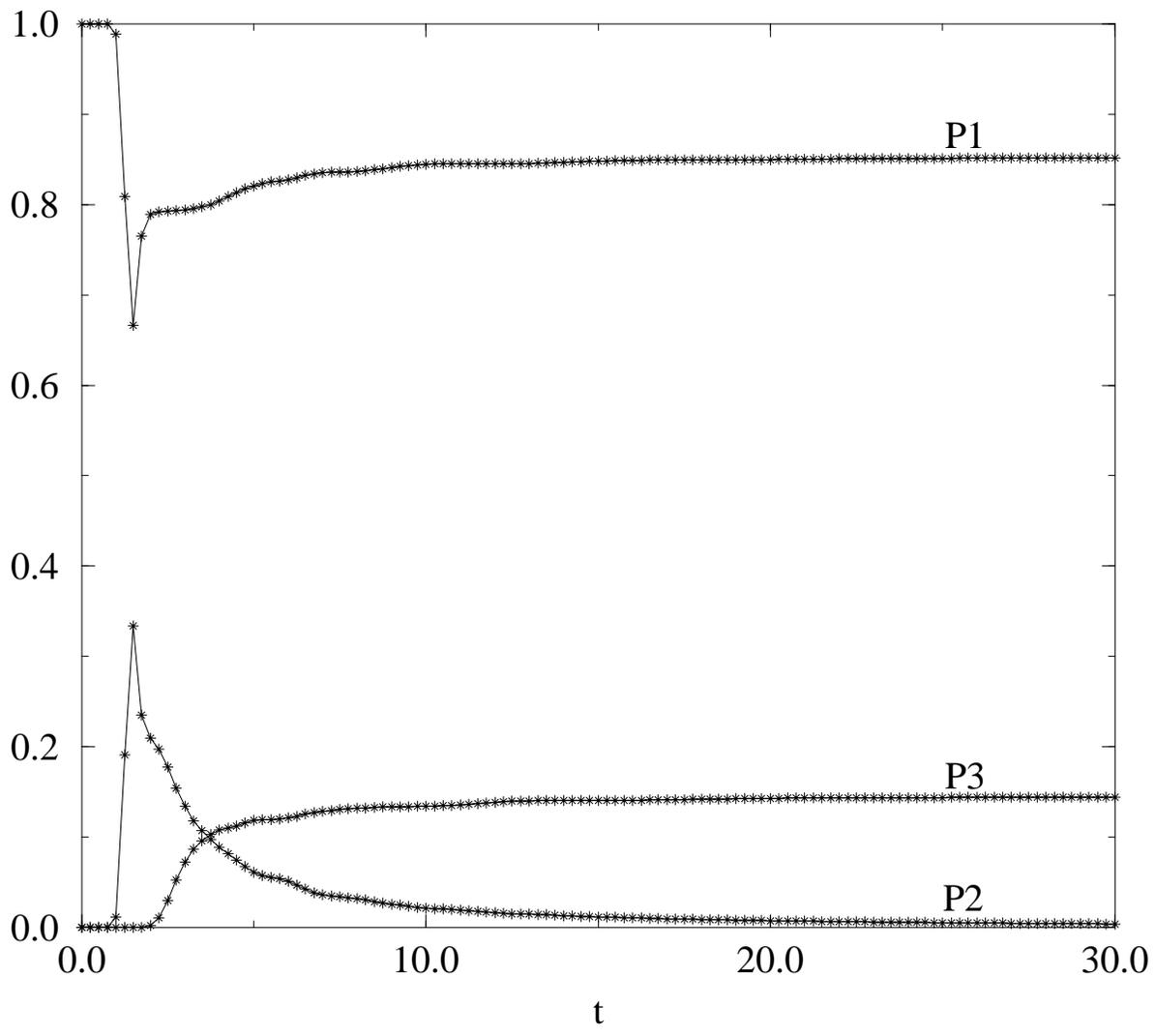
\end{document}